\documentclass[aps,prl,reprint,superscriptaddress]{revtex4-1}

\usepackage{graphicx}

\begin{document}


\title{Imaging current-induced switching of antiferromagnetic domains in CuMnAs}


\author{M.~J.~Grzybowski}
\affiliation{School of Physics and Astronomy, University of Nottingham, University Park, Nottingham NG7 2RD, United Kingdom}
\affiliation{Institute of Physics, Polish Academy of Sciences, Aleja Lotnikow 32/46, PL-02668 Warsaw, Poland}
\author{P.~Wadley}
\affiliation{School of Physics and Astronomy, University of Nottingham, University Park, Nottingham NG7 2RD, United Kingdom}
\author{K.~W.~Edmonds}
\affiliation{School of Physics and Astronomy, University of Nottingham, University Park, Nottingham NG7 2RD, United Kingdom}
\author{R.~Beardsley}
\affiliation{School of Physics and Astronomy, University of Nottingham, University Park, Nottingham NG7 2RD, United Kingdom}
\author{V.~Hills}
\affiliation{School of Physics and Astronomy, University of Nottingham, University Park, Nottingham NG7 2RD, United Kingdom}
\author{R.~P.~Campion}
\affiliation{School of Physics and Astronomy, University of Nottingham, University Park, Nottingham NG7 2RD, United Kingdom}
\author{B.~L.~Gallagher}
\affiliation{School of Physics and Astronomy, University of Nottingham, University Park, Nottingham NG7 2RD, United Kingdom}
\author{J.~S.~Chauhan}
\affiliation{School of Physics and Astronomy, University of Nottingham, University Park, Nottingham NG7 2RD, United Kingdom}
\author{V.~Novak}
\affiliation{Institute of Physics ASCR, v.v.i., Cukrovarnicka 10, 162 53 Praha 6, Czech Republic}
\author{T.~Jungwirth}
\affiliation{Institute of Physics ASCR, v.v.i., Cukrovarnicka 10, 162 53 Praha 6, Czech Republic}
\affiliation{School of Physics and Astronomy, University of Nottingham, University Park, Nottingham NG7 2RD, United Kingdom}
\author{F.~Maccherozzi}
\affiliation{Diamond Light Source, Chilton, Didcot, Oxfordshire, OX11 0DE, United Kingdom}
\author{S.~S.~Dhesi}
\affiliation{Diamond Light Source, Chilton, Didcot, Oxfordshire, OX11 0DE, United Kingdom}


\date{\today}

\begin{abstract}
The magnetic order in antiferromagnetic materials is hard to control with external magnetic fields. Using X-ray Magnetic Linear Dichroism microscopy, we show that staggered effective fields generated by electrical current can induce modification of the antiferromagnetic domain structure in microdevices fabricated from a tetragonal CuMnAs thin film. A clear correlation between the average domain orientation and the anisotropy of the electrical resistance is demonstrated, with both showing reproducible switching in response to orthogonally applied current pulses. However, the behavior is inhomogeneous at the submicron level, highlighting the complex nature of the switching process in multi-domain antiferromagnetic films. 
\end{abstract}


\maketitle

Antiferromagnetic (AF) materials are of increasing interest both for fundamental physics and applications. Recent advances in detecting and manipulating AF order electrically have opened up new prospects for these materials in basic and applied spintronics research \cite{Jungwirth16, Urazhdin07, Gomonay10, Hals11, Cheng14, Zelezny14, Wadley16}. Of particular interest is the N\'{e}el order spin-orbit torque (NSOT) \cite{Zelezny14}, recently demonstrated in the collinear AF CuMnAs \cite{Wadley16}, where a current-induced local spin polarization can exert a rotation of the magnetic sublattices. NSOT is closely analogous to the spin-orbit torque in ferromagnets with broken inversion symmetry, in which electrical currents induce effective magnetic fields that can be used to switch the magnetization direction \cite{Chernyshov09, Miron10}. The tetragonal CuMnAs lattice \cite{Wadley13} is inversion symmetric, so that zero \textit{net} spin polarization is generated by a uniform electric current. However, its Mn spin sublattices form inversion partners, resulting in \textit{local} effective fields of opposite sign on the AF-coupled Mn sites \cite{Zelezny14, Zheng14}. These staggered current-induced fields can be large enough to cause a non-volatile rotation of the AF spin axis \cite{Wadley16}.

Current-induced rotations of AF moments can be detected electrically using anisotropic magnetoresistance (AMR), a dependence on the relative orientation of the current and spin axes which is present in both ferromagnetic and AF materials \cite{McGuire75, Park11, Marti14, Kriegner16}. This provides only spatially averaged information over the probed area of the device, which may be several microns or larger. PhotoEmission Electron Microscopy (PEEM), with contrast enabled by X-ray Magnetic Linear Dichroism (XMLD), provides direct imaging of AF domains with better than 100~nm resolution \cite{Scholl00}. Based on differences in absorption of x-rays with linear polarization, XMLD-PEEM has offered valuable insights into the microscopic magnetic properties of AF films \cite{Bezencenet11} and ferromagnet / AF interfaces \cite{Nolting00, Folven12}. The measured intensity varies as $I_0 + I_2$cos$^2\alpha$, where $\alpha$ is the angle between the x-ray polarization and the spin axis \cite{Alders98}, so is equally present for AF and FM materials, similar to AMR. The XMLD amplitude given by $I_2$ also depends on the orientation of the x-ray polarization with respect to the crystalline axes \cite{Czekaj06, Freeman06}, and the signal is sensitive to domains within the top few nanometers of the surface.

\begin{figure}
\includegraphics[trim= 20 350 50 10, clip, width=0.75\textwidth]{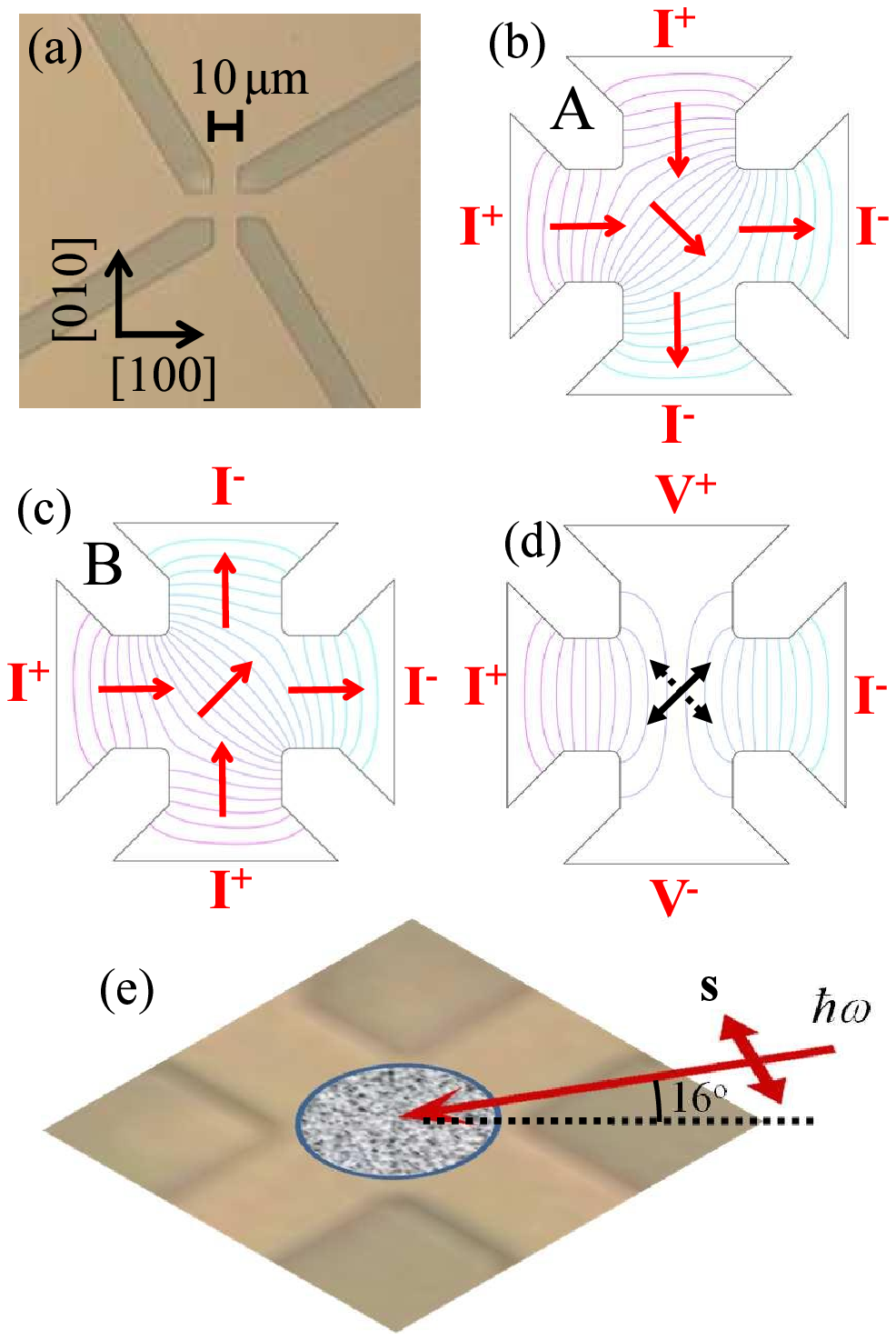}
\caption{\label{fig1}(a) Optical micrograph of the CuMnAs cross device. (b,c) The two current pulse geometries used. The arrows represent the current directions and the contours are the electrostatic potential distribution obtained by finite-element calculation. (d) Geometry used for probing the magnetic state electrically. The two magnetic states set by the current pulses, illustrated by double-headed arrows, result in opposite-in-sign transverse voltages (V$^+$ -- V$^-$) due to the AMR. (e) Geometry used for the XMLD-PEEM measurements. X-rays are incident at 16$^\circ$ to the sample surface, with polarization vector $\mathbf{s}$ in the plane of the film.}
\end{figure}

Here, we combine electrical and XMLD-PEEM measurements to demonstrate the microscopic origin of current-induced electrical switching in CuMnAs microdevices. Although the magnitude of the XMLD in semimetallic CuMnAs is significantly weaker than is typically observed in oxide antiferromagnets, we observe clear submicron AF domain structures which are systematically modified by applied current pulses, consistent with the interpretation of earlier all-electrical studies \cite{Wadley16}.

The 80~nm thick tetragonal CuMnAs film used in this study was grown by molecular beam epitaxy on a GaAs(001) substrate \cite{Wadley13}. The sample was capped with 2~nm Al to prevent oxidation. The film has a N\'{e}el temperature of 480~K and a resistivity of 160~$\mu\Omega$cm.\cite{Wadley13,Wadley15,Wadley16} Four-arm cross-shaped devices, with 10~$\mu$m wide arms oriented along the $[100]$ and $[010]$ crystal axes of the CuMnAs film (Fig.~1(a)), were prepared by photolithography and wet chemical etching. Figures 1(b) and 1(c) show finite-element calculations of the equipotentials during the application of current pulses. The pulses were applied along all four arms of the structure, either in configuration A (Fig.~1(b)) or configuration B (Fig.~1(c)), producing a net current along $[1\bar{1}0]$ or $[110]$ directions in the center of the cross. A resulting component of the spin axis along $[110]$ or $[1\bar{1}0]$ directions should then produce opposite voltages measured in the 4-probe configuration shown in Fig.~1(d), due to the transverse AMR \cite{Wadley16, McGuire75, Marti14}. Electrical current pulsing and probing were performed in-situ inside the PEEM chamber. 

\begin{figure}
\includegraphics[trim= 65 45 15 48, clip, width=0.555\textwidth]{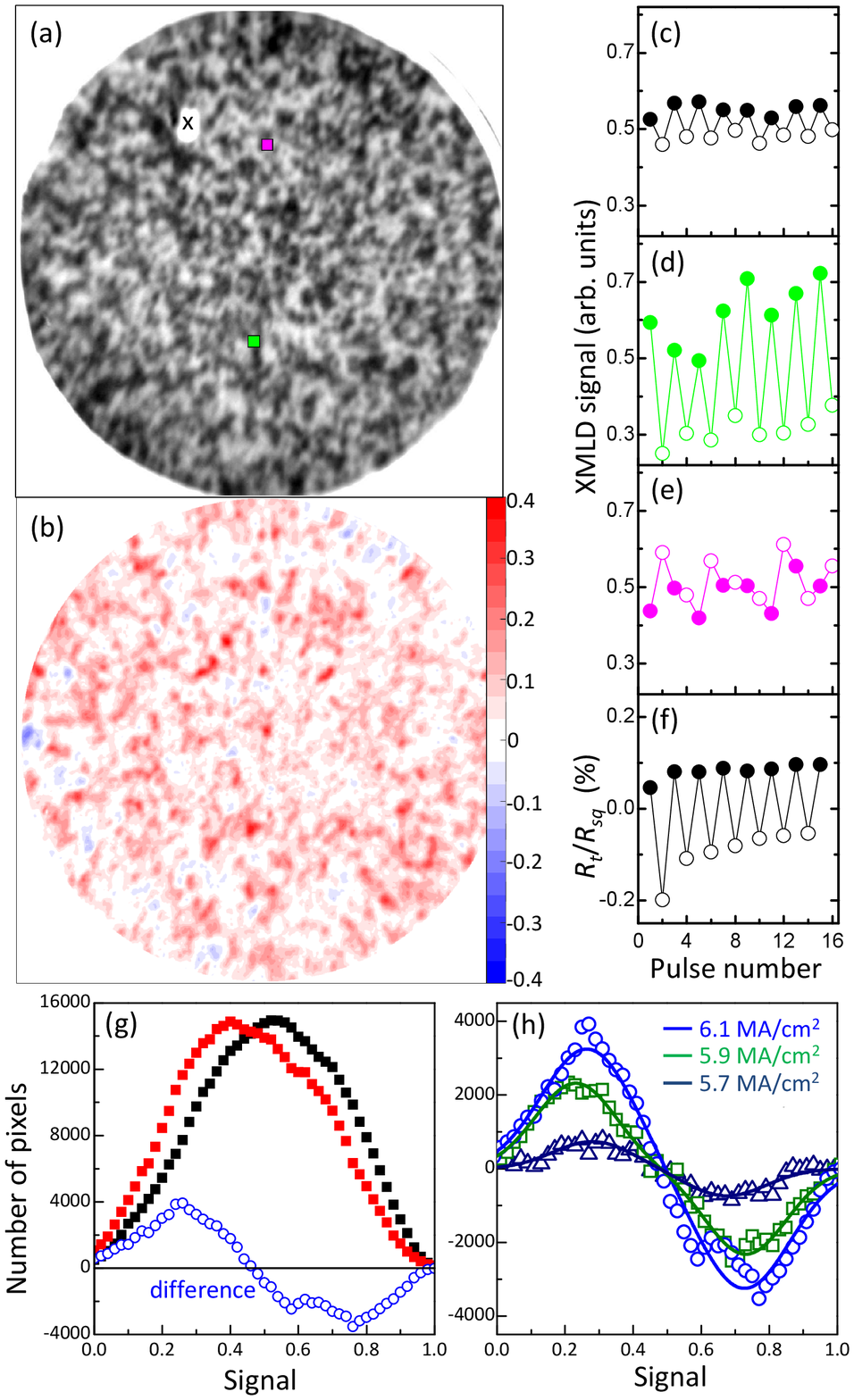}
\caption{\label{fig2}(a) XMLD-PEEM image, with 10~$\mu$m field-of-view, taken over the central section of the device. The white region marked 'x' corresponds to a defect on the device surface. (b) Difference between XMLD-PEEM images taken after applying alternate orthogonal current pulse trains of 6.1~MAcm$^{-2}$. (c) Spatially averaged XMLD signal after each pulse train. Open and filled symbols represent the two orthogonal pulse directions. (d,e) As for (c), but for the 200$\times$200~nm$^2$ regions marked by green and pink squares in (a), respectively. (f) Change in the transverse resistance following the same pulse sequence. A constant offset due to a small misalignment of voltage probes was subtracted from the transverse resistance signal. (g) XMLD intensity distribution after applying current pulses in the two configurations. The difference between the two distributions is shown by the open symbols. (h) Change in the XMLD intensity distribution for different pulse amplitudes. The points are the measured data and the lines are fits to the sum of two Gaussians.}
\end{figure}

The PEEM measurements were performed at room temperature on beamline I06 at Diamond Light Source. The x-ray beam was incident at 16$^\circ$ from the sample surface, with its polarization vector in the plane of the film along one of its $\langle110\rangle$ axes (Fig.~1(e)). XMLD contrast was obtained by taking PEEM images with x-ray energy at the Mn $L_3$ absorption edge ($E_1$) and at 0.9~eV below the edge ($E_2$). These correspond to the peak and the valley of the Mn $L_3$ XMLD spectrum, resulting in a $\approx$1\% difference in absorption between regions with spin axis parallel or perpendicular to the x-ray polarization vector \cite{Wadley15}. The x-ray absorption and XMLD spectra are shown in the Supplemental Material \cite{supple}. The XMLD-PEEM images were obtained by calculating the asymmetry, $(I(E_1) - I(E_2))/(I(E_1) + I(E_2))$, where $I(E_{1,2})$ are the measured intensities at the two energies. 

Figure 2(a) shows an XMLD-PEEM image taken from the central region of the CuMnAs device with 10~$\mu$m field of view. Sub-micron scale contrast is observed. The dependence of the contrast on the x-ray polarization and sample orientation demonstrates its predominantly magnetic origin \cite{supple}. The contrast is strongest when the polarization of the incident x-ray beam lies in the plane of the film, indicating that the antiferromagnetic moments are oriented in the film plane at varying angles with respect to the [110] axis. This is consistent with ab initio calculations \cite{Wadley15} which predict that the easy axis of tetragonal CuMnAs is in the (001) plane, with a small energy difference between [100] and [110] easy axes. The light / dark regions correspond to domains with spin axis parallel / perpendicular to the x-ray polarization, respectively. 

The difference between XMLD-PEEM images obtained after applying current pulse trains in orthogonal directions, with each train consisting of three pulses of amplitude $6.1\times10^6$~Acm$^{-2}$ and duration 50~ms, is shown in Fig.~2(b). In total, eight orthogonal pairs of pulse trains were applied. The XMLD-PEEM images obtained after each pulse train are normalized such that the brightest and darkest regions (neglecting the defect marked 'x' in Fig.~2(a)) correspond to an intensity of 1 and 0, respectively. The XMLD signal after each pulse train, averaged over the whole image, is shown in Fig.~2(c). A movie showing the XMLD-PEEM image after each pulse train is included in the Supplemental Material.

On average, the XMLD signal alternates with each successive pulse train. The sign of the XMLD difference indicates a rotation of the AF moments towards a direction perpendicular  to the current, consistent with the NSOT mechanism of current-induced switching \cite{Zelezny14, Wadley16}.  However, as is seen in Fig. 2(b), the rotation shows a submicron scale non-uniformity across the image, with some isolated regions showing a much larger change than the average, while other regions appear to switch in the opposite direction. This is confirmed in Figs.~2(d) and 2(e), which are the average XMLD signal over the 200~nm$\times$200~nm regions marked with colored squares in Fig. 2(a). Figure~2(f) shows the transverse AMR signal, $R_t$, after each current pulse train, expressed as a percentage of the sheet resistance $R_{sq}$. This shows the same switching behavior as the average XMLD.

Figure~2(g) shows distribution curves of the XMLD-PEEM signal across the device center, after applying the $J$ = $6.1\times10^6$~Acm$^{-2}$ pulse trains in the two orthogonal directions. A clear shift in the peak of the distribution is observed, consistent with an increase in the population of AF domains whose spin axis lies perpendicular to the direction of the electrical current pulses. The difference between the two distributions decreases with decreasing current pulse amplitude (Fig.~2(h)) while keeping a constant shape. 

For a purely magnetic signal with no preferential orientation of the magnetic moments in the plane of the film, the XMLD intensity distribution histogram is expected to be sharply peaked at the values corresponding to the XMLD contrast for spins at 0$^\circ$ and 90$^\circ$ to the x-ray polarization (\textit{i.e.}, the turning points of the cos$^2\alpha$ function). Due to non-magnetic contributions to the signal (\textit{i.e.}, instrument resolution and noise), the peaks are substantially broadened resulting in the distributions seen in Fig. 2(g). The difference between the two distributions after orthogonal current pulses can be fitted as the sum of two Gaussians (Fig. 2(h)), with the separation of the Gaussian centers corresponding approximately to the amplitude $I_2$ of the XMLD. From this we estimate that $I_2\approx0.4$ on the normalized scales shown in Fig.~2. The average change in the XMLD after the $J$ = $6.1\times10^6$~Acm$^{-2}$ current pulse trains is around 0.07$\pm$0.01 (Fig.~2(c)), which corresponds to an average rotation of the magnetic moments of around 10$^\circ$. Meanwhile for small isolated regions such as the one described by Fig.~2(d), the rotation of the moments approaches the full 90$^\circ$.

\begin{figure}
\includegraphics[trim= 30 345 10 10, clip, width=0.50\textwidth]{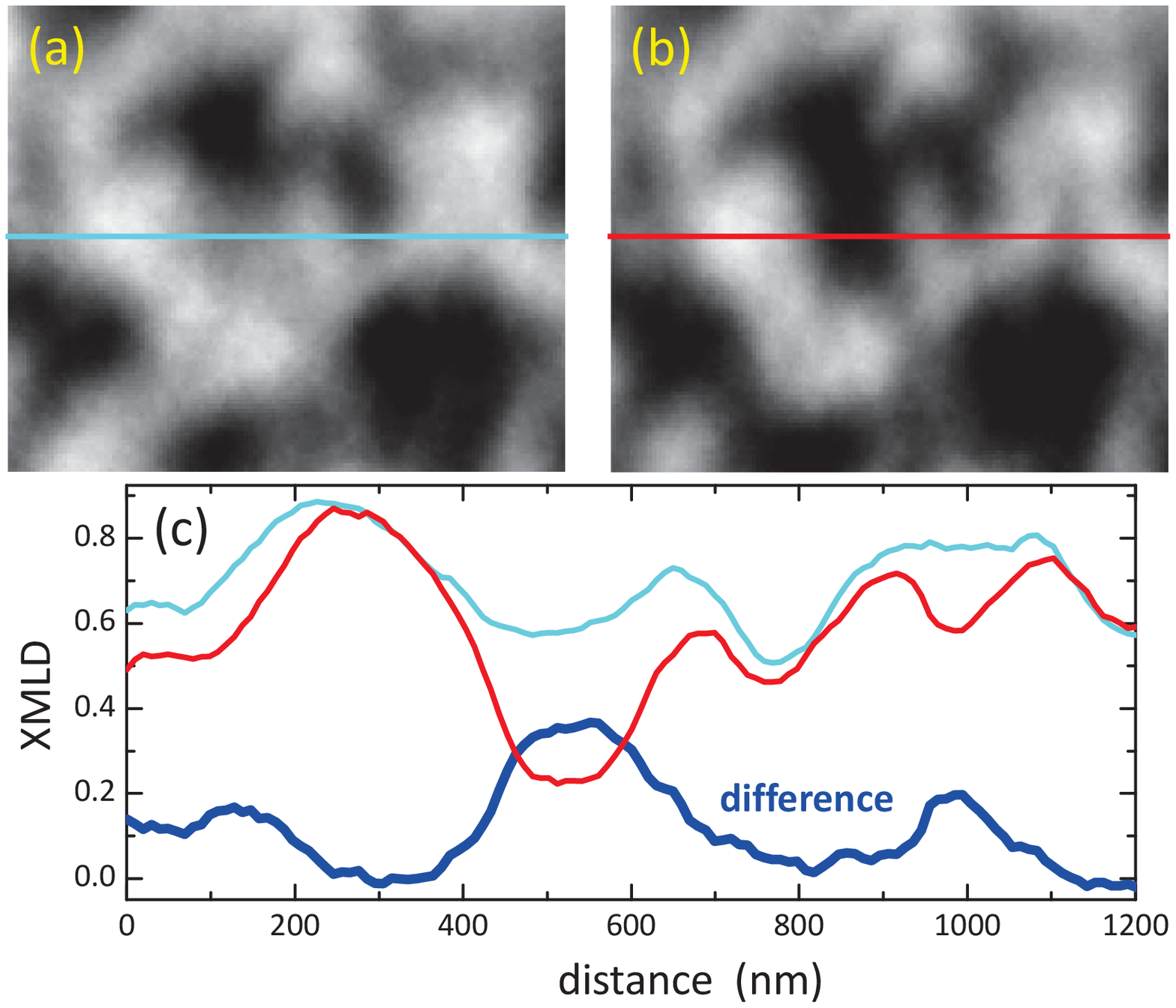}%
\caption{\label{fig3}(a,b) XMLD-PEEM contrast for the same 1.0$\times$1.2~$\mu$m region of the CuMnAs microdevice, after applying current pulses of 6.1~MAcm$^{-2}$ in the two orthogonal configurations shown in Fig. 1. (c) Linescans through the centers of the images in (a,b), and their difference.}
\end{figure}

To further demonstrate the inhomogeneous nature of the current-induced switching, Figs.~3(a) and 3(b) show expanded views of a 1.2$\mu$m$\times1\mu$m region close to the center of the device, after applying current pulses in orthogonal directions. The movement of AF domains can be observed in several locations across the image, while other locations show no change in contrast. This is also seen in Fig. 3(c), which shows linescans through the center of the images in Fig.3(a,b). The large changes are observed in isolated regions of typical width $\approx$100-200~nm. Calculations of NSOT switching in disorder-free AF systems have previously shown coherent rotation and domain wall propagation at very short timescales \cite{Zelezny14, Gomonay16, Roy16}. However, the observed inhomogeneous texture and localized switching is not unexpected given that AF domains are considered to be dominated by magnetoelastic deformations and elastic defects such as disclinations and dislocations \cite{Gomonay07}. This situation is further compounded at the surface which may show distinct magnetoelastic properties to the bulk \cite{Gomonay02}. The AF domain structure and distribution of elastic fields are most likely seeded by the substrate during the growth process. The local switching is then a natural consequence of the elastic strain within the film. 

\begin{figure}
\includegraphics[trim= 20 412 0 12, clip, width=0.5\textwidth]{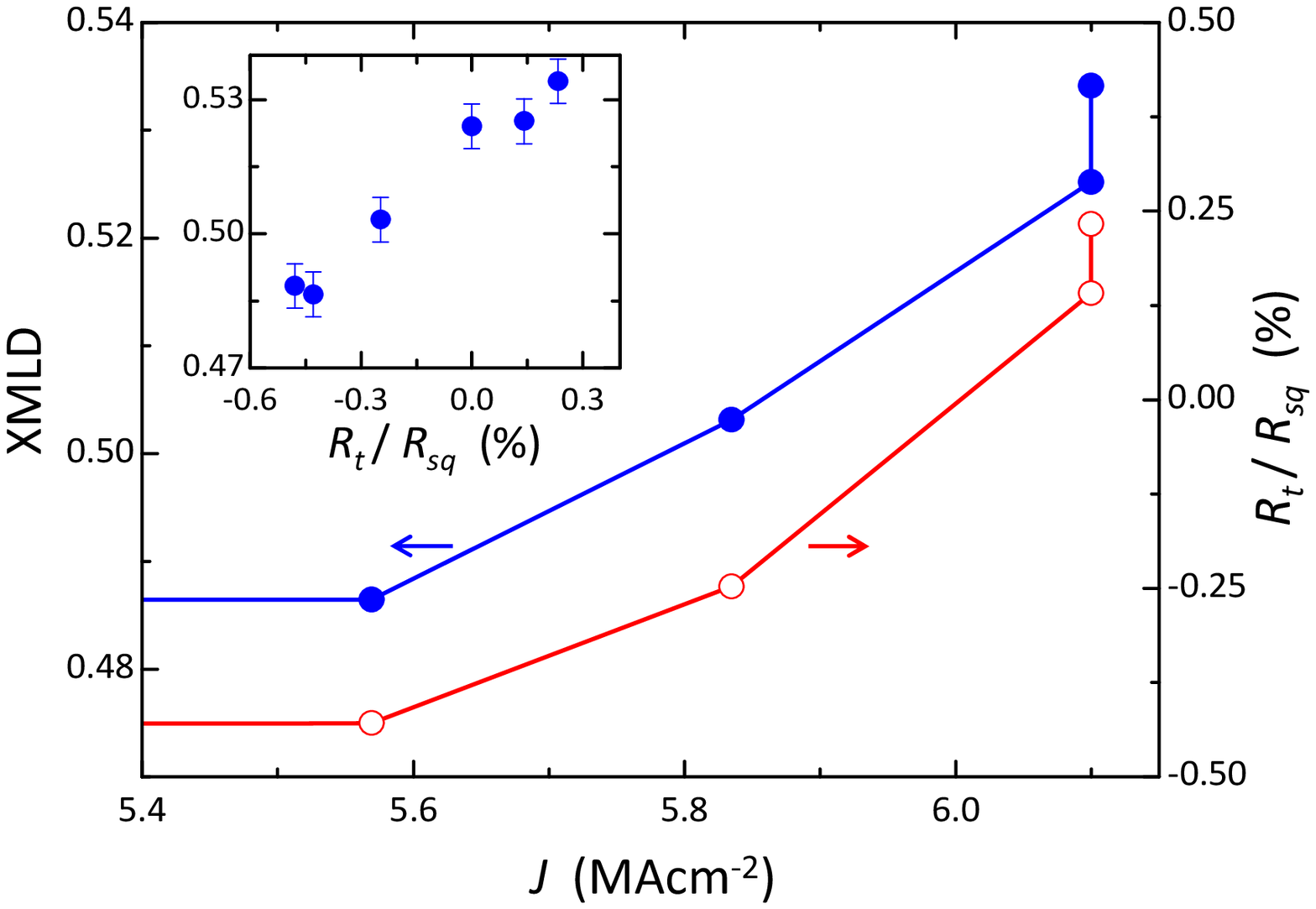}
\caption{\label{fig4}(a) Mean XMLD signal over the center of the device (filled symbols, left axis) and transverse resistance (open symbols, right axis) versus current pulse density. The device was first set by applying current pulses of 6.1~MAcm$^{-2}$ in the orthogonal direction. Inset: mean XMLD versus transverse resistance.}
\end{figure}

To establish the correlation between the XMLD signal and the anisotropic electrical resistance, we investigated their dependence on the amplitude of the current pulse. The sample was initially set by applying three $J$=$6.1\times10^6$~Acm$^{-2}$ current pulses in configuration A (Fig.~1(b)). Then, single pulses of increasing amplitude were applied in the orthogonal configuration B, collecting an XMLD image and a transverse resistance measurement after each pulse. As a final step, a train of three $J$=$6.1\times10^6$~Acm$^{-2}$ pulses was applied in configuration B. Figure 4 shows the mean XMLD over the center of the device and the transverse resistance recorded after each step. The two measurements show similar behavior, as expected since both the AMR and the XMLD follow a similar dependence on the relative orientations of spin and reference (current or polarization) axes. The inset shows that the mean XMLD varies linearly with the transverse resistance. This is therefore confirmation that the observed transverse resistance changes induced by the current pulses are due to the reorientation of AF domains. Moreover, since the observed AMR corresponds to only a small average rotation of the AF moments, a much larger electrical signal may be anticipated if full reorientation could be achieved. By extrapolating the measured signals to the maximum XMLD amplitude of $I_2\approx0.4$, a maximum AMR of (3$\pm$1)\% in the antiferromagnetic CuMnAs film is estimated for full reorientation.

In summary, our results show directly the switching of antiferromagnetic moments due to current-induced N\'{e}el-order spin-orbit torque, providing confirmation of recent theories and electrical probes. The staggered effective magnetic field generated by the current provides a means of manipulating nanoscale AF domains and domain walls, opening routes to new memory technologies and new research into the dynamics of AF coupled spins. Future work with newly available aberration-corrected PEEM instruments could highlight the role played by AF domain walls and interdomain, as well as intradomain, inhomogeneity.

\begin{acknowledgments}
We thank Diamond Light Source for the allocation of beamtime under proposal number SI12504. We acknowledge support from the University of Nottingham EPSRC Impact Acceleration Account grant no. EP/K503800/1, the EU 7th Framework Programme under the project REGPOT-CT-2013-316014 (EAgLE), the EU ERC Advanced Grant No. 268066, the Ministry of Education of the Czech Republic Grant No. LM2011026, and the Grant Agency of the Czech Republic Grant no. 14-37427.
\end{acknowledgments}

\bibliography{XPEEMrefs}

\end{document}